\RequirePackage{snapshot}
\documentclass[conference]{IEEEtran}
\usepackage{flushend}

\usepackage{ifpdf}

\usepackage{cite}

\ifCLASSINFOpdf
  \usepackage[pdftex]{graphicx}
\else
  \usepackage[dvips]{graphicx}
  \DeclareGraphicsExtensions{.eps}
\fi
\graphicspath{{img/}} 
\usepackage{epsfig}

\usepackage{stfloats}

\usepackage{slashbox}

\usepackage[utf8x]{inputenc}
\usepackage{xcolor}
\usepackage{color,soul}
\usepackage{todonotes}

\begin{document}

%
\title{Explosive pulsed power to drive a vacuum tube}

\author{\IEEEauthorblockN{
A.~Gurinovich}
\IEEEauthorblockA{Research Institute for Nuclear Problems, Belarusian State University, Minsk, Belarus\\
Email: gur@inp.bsu.by}}


%


\maketitle

\begin{abstract}
Development of high-power pulsed radiation sources in any frequency range requires both generation of high power to drive the source and increasing the efficiency of supplied power to radiated electromagnetic field conversion. The former implies generation of high power (that is equal to high-voltage and high-current) pulses. The latter means use of an electron beam moving in vacuum to produce intense radiation: high electron beam current or high current density combined with large cross-section of interaction area are required. Explosive pulsed power could contribute to both of the above being capable to store and deliver much higher specific energy as compared to either dielectrics or magnetics and providing high flexibility for matching with a load by the use of a pulse forming network. 
Piecemeal matching of explosively driven power supply with the HPM producing load (vacuum tube) is described.
%
%
\end{abstract}



%
\IEEEpeerreviewmaketitle

\section{Introduction}

Explosively driven particle accelerators \cite{1,2} and high-power microwave
sources \cite{3,4} are being developed during recent decades.
High supplied power, compact size and wide range of achievable output parameters are among the
advantages of such devices. 
However, development of any explosively driven source with high repeatability of parameters is complicated  since it is fundamentally single
shot.
This paper considers a consistent process of design,  development and testing of an explosively driven high power microwave pulsed power system composed of a helical flux compression generator (FCG)
seeded from a capacitor bank and feeding a power conditioning system, which
includes a pulsed transformer and a fast opening
electro-explosive switch with inductive energy storage, and a vacuum tube capable to produce
high-power microwaves. 
Development approach ensuring optimal operation of the system as a whole includes subsystems analysis and successive testing of subsystems in different combinations.

The paper is organized as follows: Section II presents system overview and subsystems analysis. Section III includes key parameters of subsystems. Section IV describes development approach and  illustrates each step by discussion of experimental results. Optimization opportunities are discussed in Conclusion along with concluding remarks. 

\section{System overview and subsystems analysis}

An explosively driven high power microwave tube (referred to hereinbelow as ''system'') is composed of five subsystems as follows: a seed
source, which requires a charger and a trigger to synchronize explosive and electric processes, a helical flux compression generator (FCG), a transformer and an exploding wire array (EWA) both serving for conditioning power delivered by FCG to drive a vacuum tube.

Design options for each subsystem are thoroughly described in numerous research papers and books (see, for example \cite{1,2,3,4,A1,A2,A3,A4,A5}).
%

\begin{figure}[htb]
	\centering 
	\includegraphics[width=8 cm]{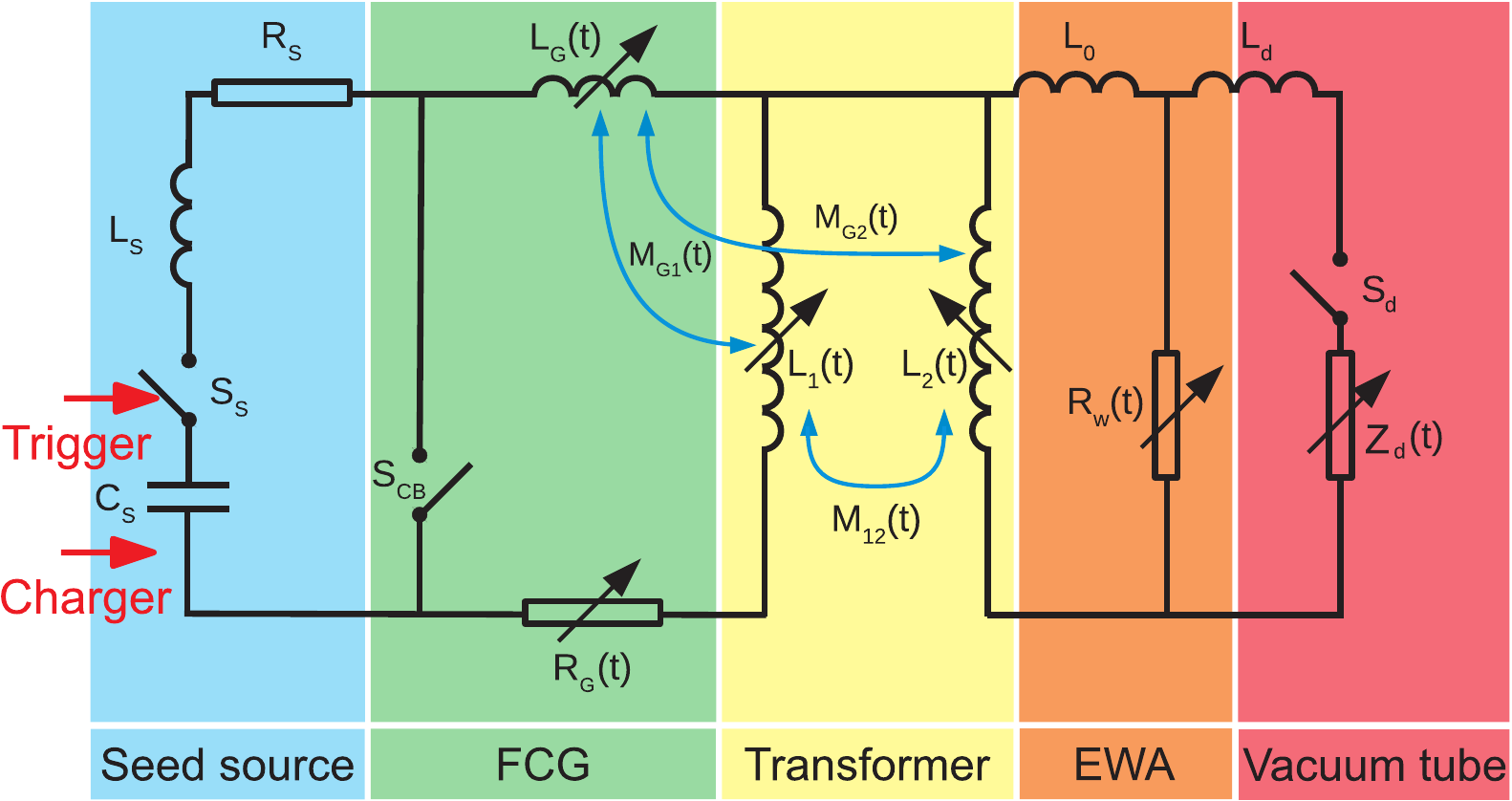}\\
	\caption{An explosively driven high power microwave source: electric circuit diagram combined with block diagram }\label{fig:circuit}
\end{figure}

\begin{figure}[htb]
	\centering 
	\includegraphics[width=8 cm]{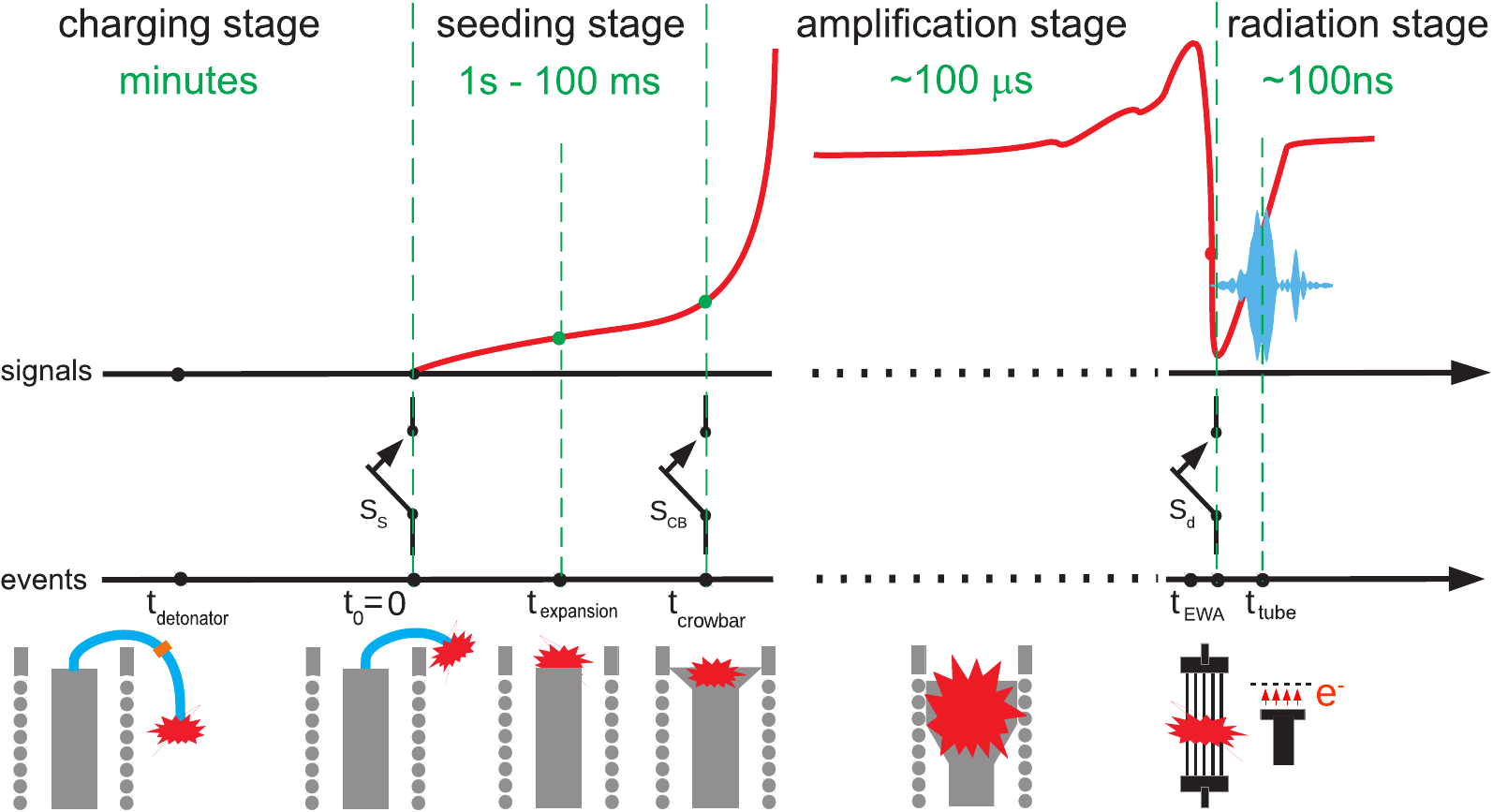}\\
	\caption{Sequence of operation stages: charging stage, seeding stage, amplification stage and radiation stage: typical time-scale for each stage is indicated in green, red curves illustrate current derivative in the circuit, switches closing instants are also shown}\label{fig:T_scale}
\end{figure}

Operation of the system as a whole can be illustrated by an electric circuit combined with block diagram and a chart showing sequence of operation stages (see
Fig.\ref{fig:circuit} and Fig.\ref{fig:T_scale}).

Brief description of events at operation stages is as follows:

 -- during charging stage a seed
source with capacitance $C_S$ is charged to operation voltage $U_S$, then a trigger serving to synchronize explosive and electric processes by closing switch $S_S$ at instant $t_0=0$ enable supplying initial current to FCG circuit at seeding stage;

-- at $t=t_{expansion}$ at seeding stage liner starts expanding inside FCG inductor until comes into contact with it at $t=t_{crowbar}$, this instant corresponds to closing switch $S_{CB}$;

-- growing current in FCG circuit and over the primary winding of transformer at amplification stage leads to growth of voltage drop over transformer secondary winding, which is then further multiplied by exploding wire array (EWA) with time-dependent resistance $R_w (t)$;

-- thus, power delivered by FCG is conditioned to be switched by spark gap $S_d$ to drive vacuum tube with impedance $Z_d (t)$ and to produce HPM at radiation stage at $t \ge t_{tube}$.

To ensure radiation output in explosively driven shots and considering different time scales for operation stages, complexity of each subsystem and  high cost of full-scale tests,  thorough matching of subsystem parameters is required.  
A matching approach, which includes 
definition of the limiting parameters for each operation stage,
 simulation and supplementary tests to ensure subsystem features and performance, is proposed and verified in successive testing. 

\section{Key requirements to subsystem parameters}

The starting point for system development is understanding of operation parameters of two constitutive subsystems, namely: microwave tube with desired output and flux compression generator capable to produce power sufficient to supply vacuum tube.

\subsection{Vacuum tube}
\label{sec:HPM}

Since driving a vacuum tube is the goal of presented analysis 
let us start from considering parameters required for its operation. 

\begin{figure}[htb]
	\centering 
	\includegraphics[width=7 cm]{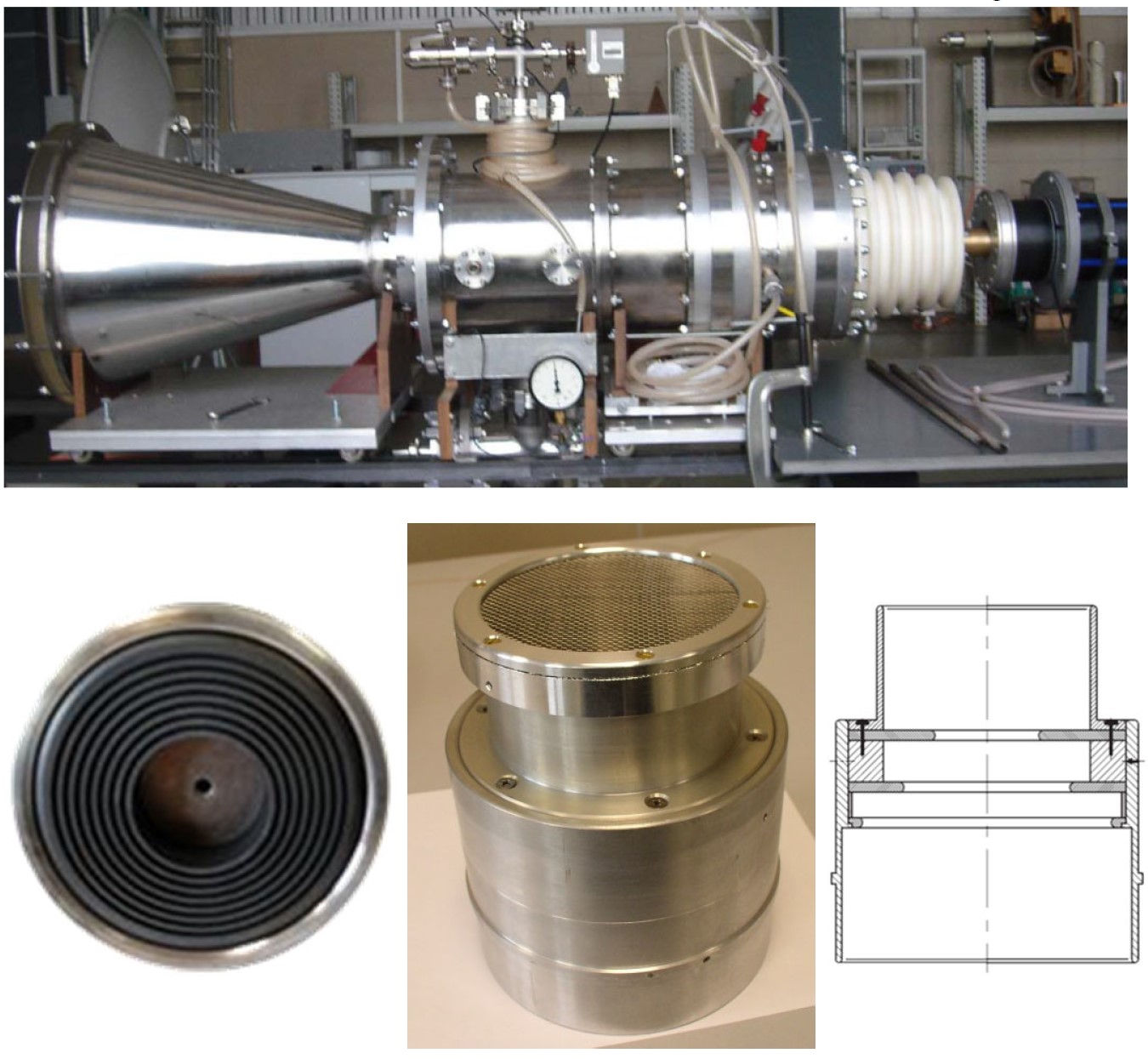}\\
	\caption{Axial vircator: general view, cathode, resonator photo and drawing \cite{A6,axial_2013,axial_2014,axial_2015,axial_2015_2}}
	\label{fig:axial}
\end{figure}

\begin{figure}[htb]
	\centering 
	\includegraphics[width=7 cm]{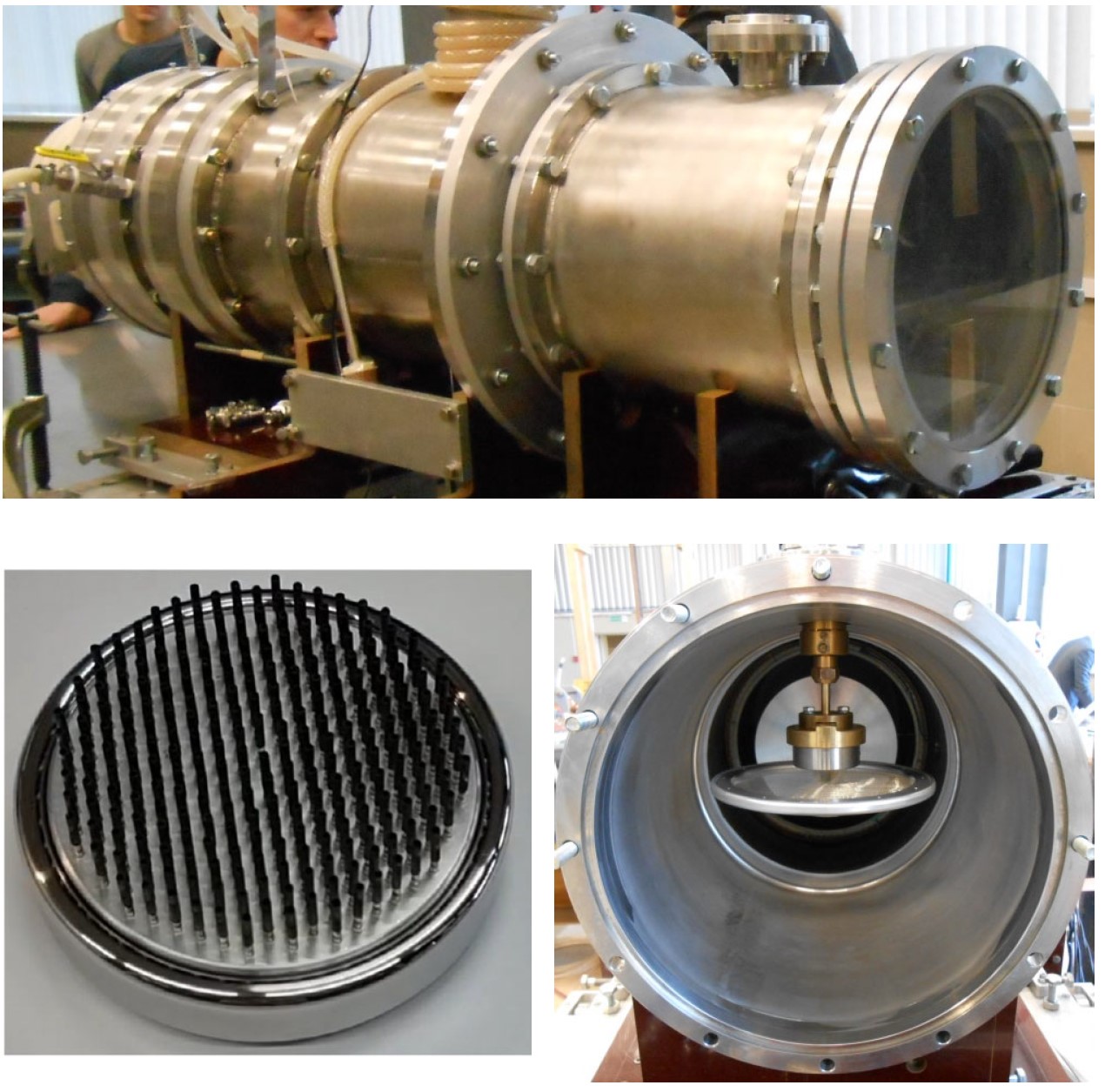}\\
	\caption{Reflex triode: general view, cathode and cavity inner view \cite{A6,reflex_2016,reflex_2017}}
	\label{fig:reflex}
\end{figure}

To reduce the number of subsystems we focused on those types of microwave tubes, which operate without guiding magnetic field. Three types of compact HPM sources operating at electron beam energy within the range 150–500 keV  were compared in \cite{A6}.  
Two of those three are selected for explosive testing, namely: axial vircator 
(see Fig.\ref{fig:axial}) and virtual cathode
oscillator in reflex triode geometry (referred as reflex triode) (see 
Fig.\ref{fig:reflex}). 
Selection criteria include capability to produce high output power and stability of output parameters (both frequency and power) at perceivable current variations. 

\begin{figure}[h!]
	\centering 
	\includegraphics[width=8 cm]{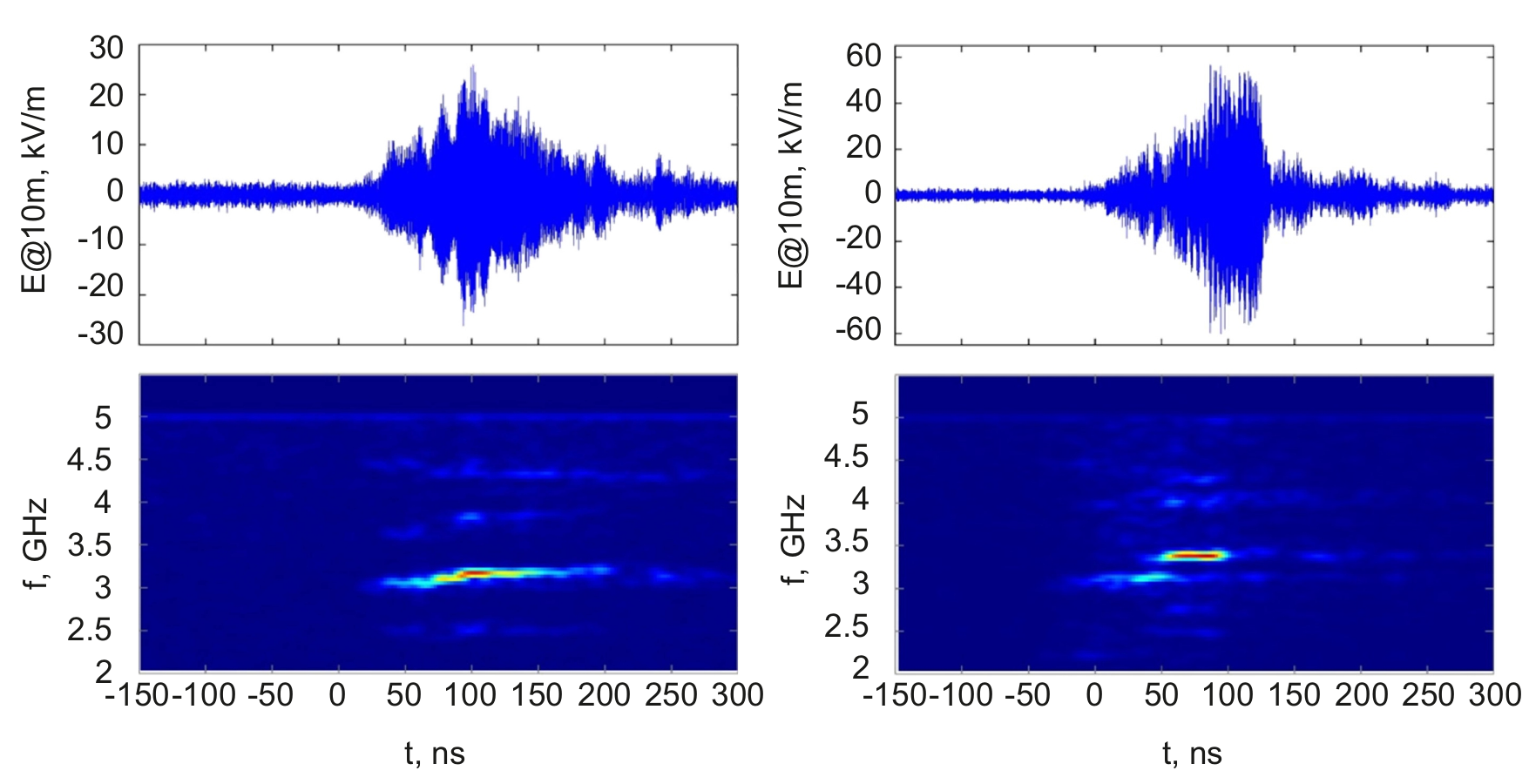}\\
	\caption{Typical time-domain waveform and its spectral contents for axial vircator (left) and for reflex triode (right)}
	\label{fig:wf_2}
\end{figure}

Summary of HPM source parameters \cite{A6} is given in Table~\ref{tab:comparison}. 
Typical time-domain waveforms and their spectral contents are shown in Fig.\ref{fig:wf_2}.

\begin{table}[h!]
	\caption{Operation parameters of axial vircator and reflex triode \cite{A6}}
	\label{tab:comparison}
	\begin{center}
		\begin{tabular}{|c|c|c|}
			\hline
			Parameter & Axial vircator  & Reflex triode \\
			\hline
			Frequency [GHz] & 3.2 $\pm$ 0.1  & 3.4 $\pm$ 0.3\\
			Epeak@10m [kV/m]
			 & 26 $\pm$ 3  & 40 $\pm$ 5\\
			Duration of microwave pulse [ns] & 60$\pm$10  & 100$\pm$50\\
			Operation voltage [kV] & 400 $\pm$ 50  & 400 $\pm$ 50\\
			Vacuum diode current [kA] & 20 $\pm$ 2  & 15 $\pm$ 4 \\
					\hline
		\end{tabular}
	\end{center}
\end{table}


Both HPM sources were tested in laboratory using a pulsed power supply comprising a 30kJ/100kV capacitor bank and an exploding wire array (EWA). Varying charging voltage of the capacitor bank, connecting inductances, length and number of wires in EWA  one could drive an HPM source with pulses, which parameters simulate those available from explosive pulsed power source (for more details see \cite{axial_2015}).
	%

According to the laboratory tests parameters required for vacuum tube driving from explosive pulsed power supply are as follows:
\begin{itemize}
	\item Input voltage 400$\pm$60~kV;
	\item Voltage pulse FWHM $>$200~ns;
	\item Voltage pulse risetime $<$70~ns; 
	\item Diode current 16$\pm$5~kA. 
\end{itemize}

\subsection{Flux compression generator}
\label{sec:FCG}

According to classification of generators for magnetic cumulation proposed in \cite{1}, to drive vacuum tube one should develop energy generating FCG: helical flux compression generator is among the options \cite{4,A1,
	Sakharov1965,
	Demidov2012,
	Selemir2019,
	EAPPC_2016_exp
}.

\begin{figure}[htb]
	\centering 
	\includegraphics[width=3 cm]{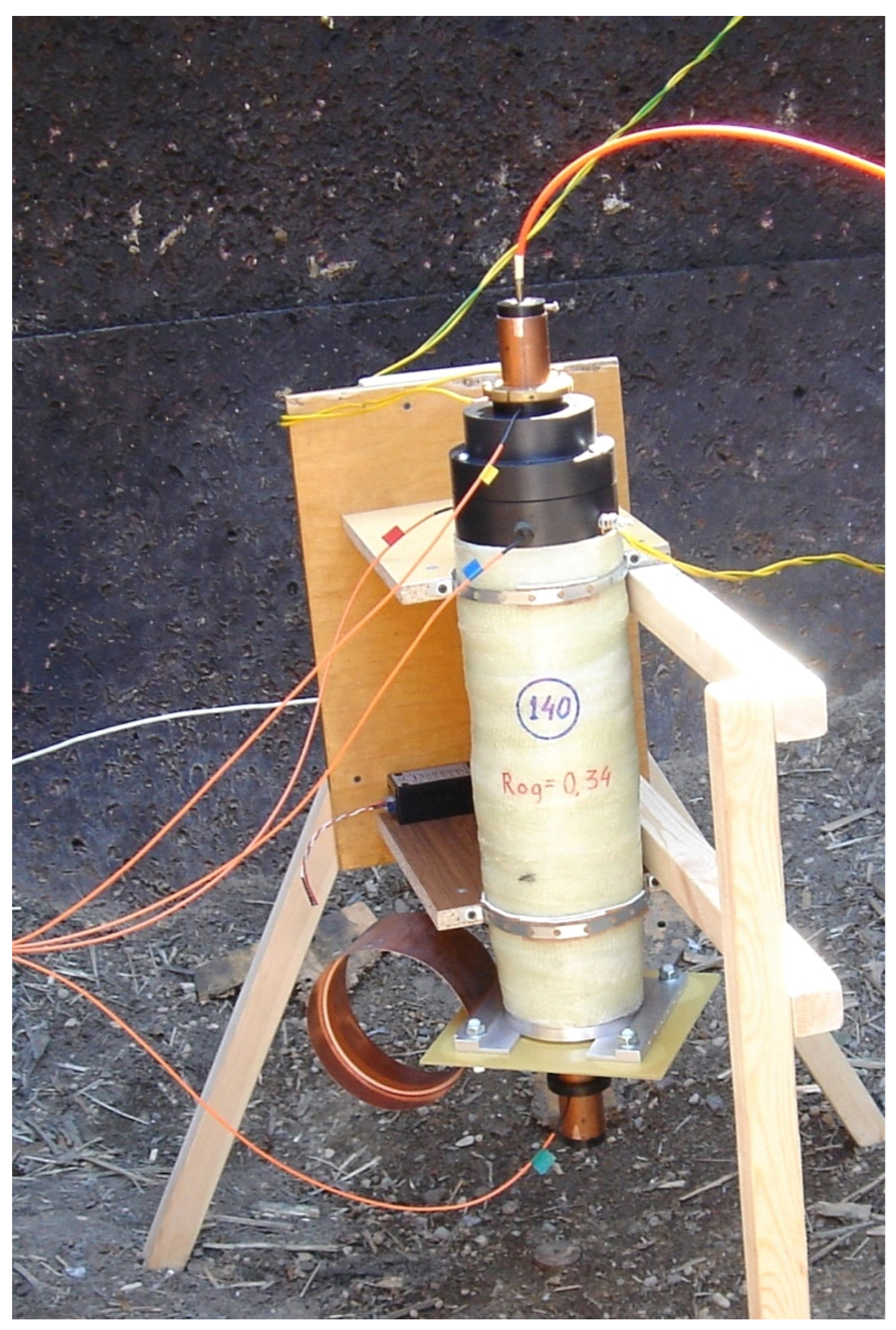}\\
	\caption{Test of FCG with purely inductive load}
	\label{fig:fcg}
\end{figure}

Single stage helical FCG with a tapered coil geometry and winding was developed to drive the system. Detonation was initiated end-on, detonation velocity approached 8mm/$\mu$s, expansion cone angle was about 11$^\circ$.
According to simulation using FCGcalc tool \cite{FCG_2017,FCG_2018} the optimal load has the effective inductance within the range 90 to 190 nH. Current derivative was requested to be as high as $8 \cdot 10^{10}$~A/s. We tested several FCGs with purely inductive loads (see Fig.\ref{fig:fcg}) from 25 to 500 nH to check simulation results. Seeding current varied from 1 to 10 kA to ensure that  FCG current fits the requirement H$<$80\,MA/m to omit nonlinear diffusion. Seeding time was required to be $<$\,400ms. 
Helical flux compression generator is directly seeded from the capacitor based stand-alone seed source including a capacitor bank, a charger, a spark-gap switch, which is controlled via optic
fiber, and a current monitor. 
Two variants of seed source were used: fast one (with capacitance 10$\mu$F) and slow that (with  capacitance 70$\mu$F).
Either positive or negative polarity can be set. Delayed ignition is used to synchronize FCG explosion and seed
current maximum: capacitor bank
discharge to FCG circuit is controlled by spark-gap closing.

Operation of FCGs of different designs with 150~nH load was described in \cite{reflex_2017}, summary of the obtained parameters is illustrated by Table~\ref{table:fcg}.
\begin{table}[h!]
\caption{Summary of FCG operation  with 150~nH load \cite{reflex_2017}}
	\label{table:fcg}
	\begin{center}
		\begin{tabular}{|l|c|c|c|}
			\hline
			FCG model &  \#1 &  \#2 & \#3  \\
			\hline
			Liner outer diameter, mm
			&  41 & 41 & 57  \\
			Max/min coil diameter, mm & 90/84  & 104/84 & 122/84   \\
			Inductor coil length, mm & 425  & 425 & 490   \\
			Number of winding sections & 7  & 6 & 6   \\
			Initial inductance, $\mu$H@10kHz
			 & 400  & 225 & 650   \\
			I$_{CB}$,~kA
			 & 1.2  & 4.0 & 2.0  \\
			Max. current derivative, A/s
			 & 8.2$\cdot$ 10$^{10}$  & 9.3$\cdot$ 10$^{10}$ & 1.1 $\cdot$ 10$^{11}$   \\
			Max. current, kA
			& 740  & 890 & 920   
			\\
			\hline
		\end{tabular}
	\end{center}
\end{table}





Preliminary tests enabled to ensure the following parameters for FCG operation with 10\% tolerance in any case when identical setting are used:
\begin{itemize}
	\item Maximal voltage about 10~kV over effective load 90-150~nH;
	\item Maximal FCG current derivative $\sim 1\cdot 10^{11}$ A/s;
	\item Maximal FCG current $<$1~MA;
	\item Power delivered to the load $\sim$10~GW.
\end{itemize}

\subsection{Transformer}
\label{sec:trans}

Initial requirements to key parameters of a transformer and its connections are as follows:
\begin{itemize}
	\item Coaxial design;
	\item Inductance of primary winding: $300 \pm 50$~nH;
	\item  Total inductance in secondary circuit $<15~\mu$H;
	\item  Mutual inductance of primary and secondary windings $> 1.25 \mu$H;
	\item  Mutual inductance of primary winding and FCG $< 500$~nH;
	\item Mutual inductance of secondary winding and FCG $< 3\mu$H;
	\item Maximal voltage drop between primary and secondary windings ~450 kV.
\end{itemize}

\begin{figure}[!h]
	\centering 
	\includegraphics[width=6 cm]{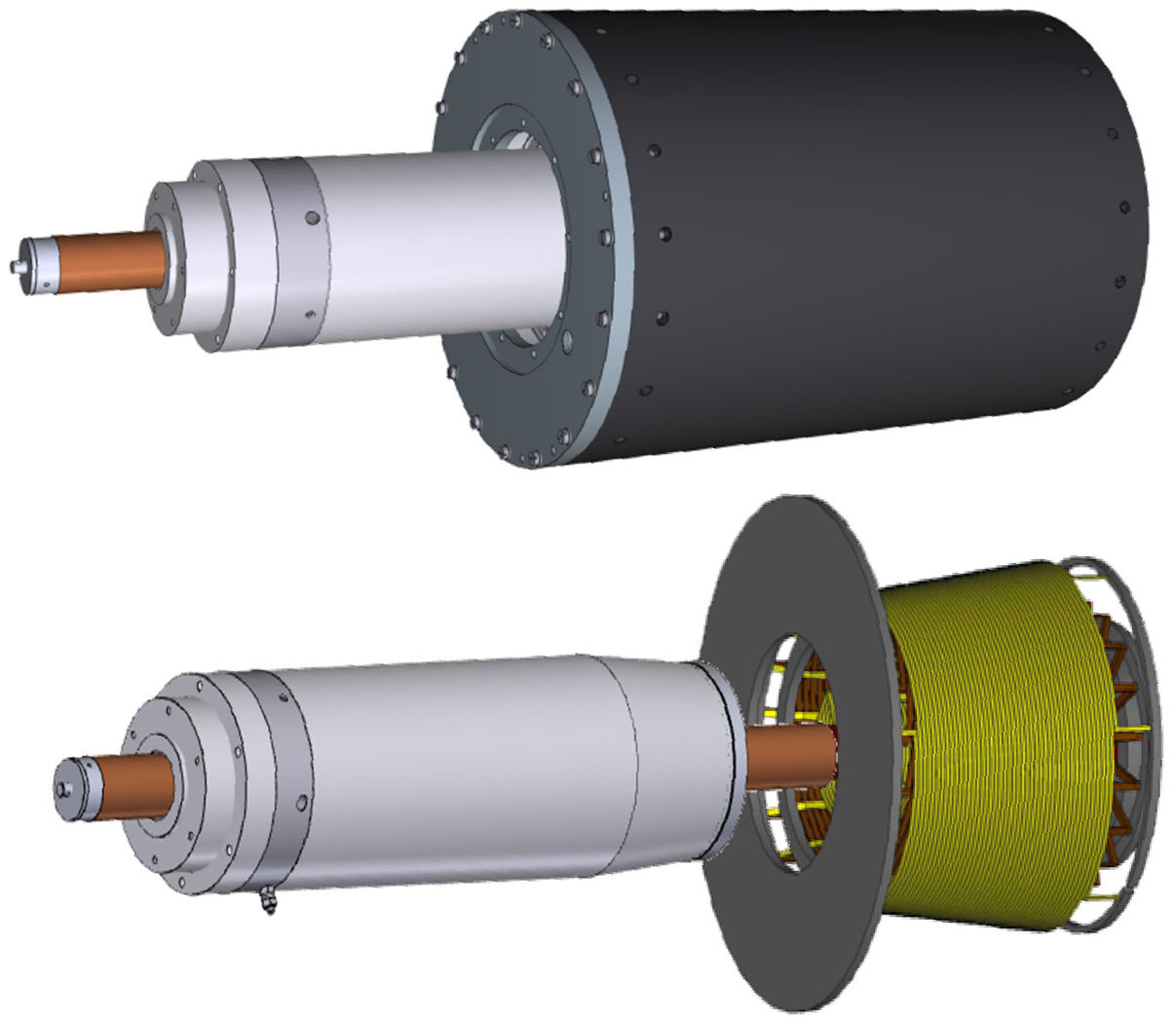}\\
	\caption{Transformer design options for different coupling of transformer windings with FCG}
	\label{fig:trans_design}
\end{figure}

Windings of a transformer in coaxial design had conical tapered shape and were manufactured using the technology similar to that used for FCG inductor coil.
Tapering was used to increase the distance between the windings in the areas with the highest voltage drop across.
Coupling of transformer windings with FCG was analyzed for different design options (see Fig.\ref{fig:trans_design}).
Inductances of all connection lines were considered, autotransformer was analyzed.
Application of either transformer oil or pressurized gas (nitrogen or SF$_6$ or mixture of two) for insulation was implied.

Each developed transformer was tested for HV durability at 450\,kV\@100ns. Driving transformer from FCG  with a combined resistive-inductive load (water-filled resistor) enables measuring currents over the transformer secondary winding in order to define initial parameters for EWA design. 

\subsection{Exploding wire array}
\label{sec:EWA}

\begin{figure}[htb]
	\centering 
	\includegraphics[width=6 cm]{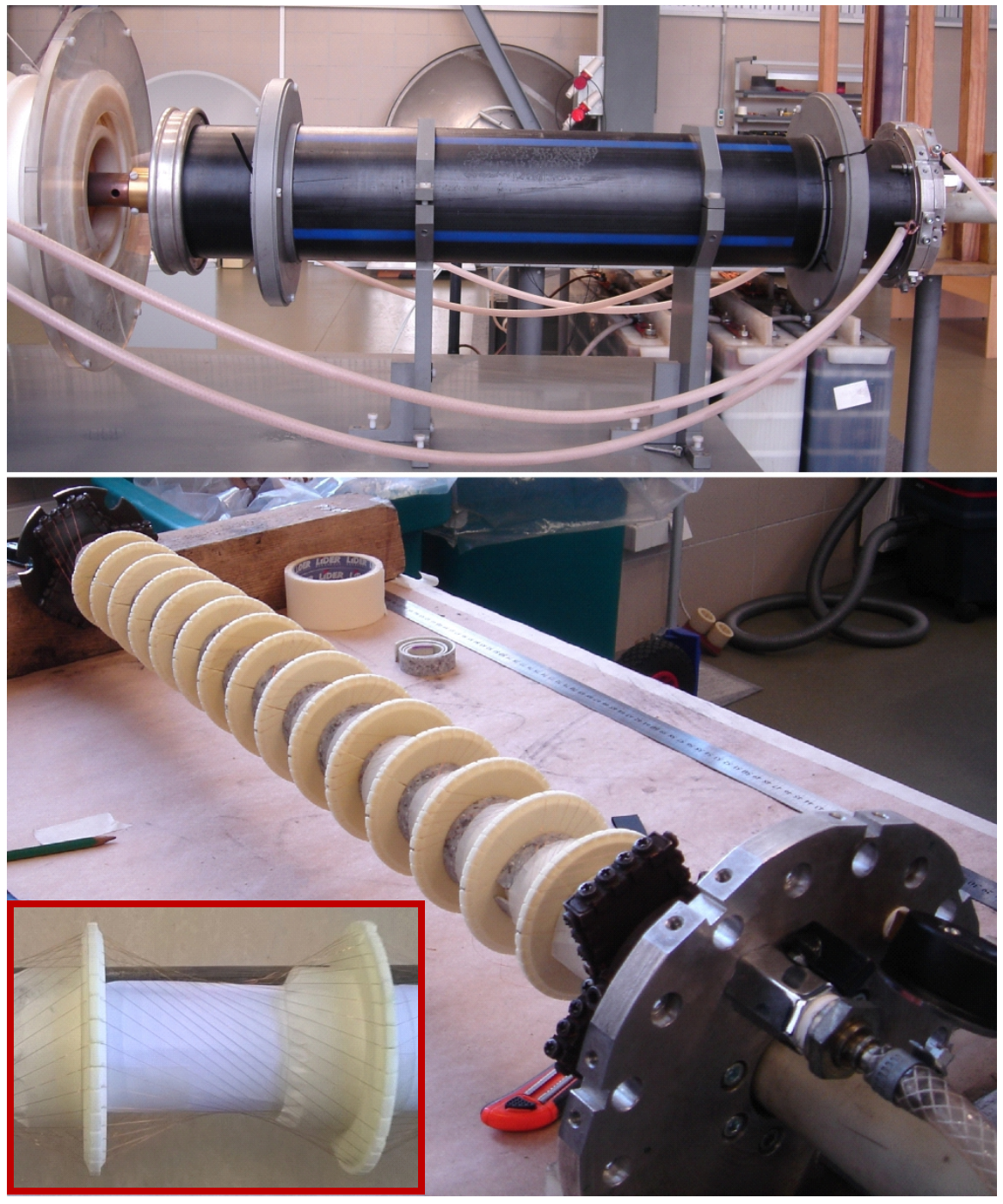}\\
	\caption{Exploding wire array can be used in either straight or zigzag packing in the case designed for filling with pressurized gas}
	\label{fig:wires}
\end{figure}

Exploding wire array (EWA) was tested in the laboratory (Fig.\ref{fig:wires}) with the use of
the same pulsed power supply we used for HPM source (see section \ref{sec:HPM})
to obtain  the length and the number of wires in EWA for the value of current obtained in the secondary circuit in transformer tests. 
HV durability of exploding wire array case was also tested up to 600\,kV\@100ns, HV strength was ensured by the use of insulating gas (nitrogen or SF$_6$ or mixture of two) at up to 0.5~MPa pressure. 
Up to 36 parallel connected oxygen-free
high-conductivity (OFHC) 99.99$\%$ purity copper wires 100~$\mu$m
in diameter with length varied from 0.3m to 0.75m for straight wires and from 0.5m to 1m for zigzag packing, which enables to reduce the overall length of EWA, were used in laboratory tests to obtain initial data for further matching. Comparison of operation of EWA comprising straight wires with that with zigzag packing at different gas pressure values enabled to fix the working pressure range ensuring identical output for both EWA designs.

\section{Matching approach and experimental results}

After laboratory testing at subsystem-level, matching of all parameters is required before the full-scale explosive system test. 
EWA parameters can be fitted by varying the number of wires and their length and all the changes can be made without subsystem re-design. Another matching parameter is the seed source voltage(current at crowbar closing instant $I_{CB}$).

\begin{figure}[htb]
	\centering 
	\includegraphics[width=8 cm]{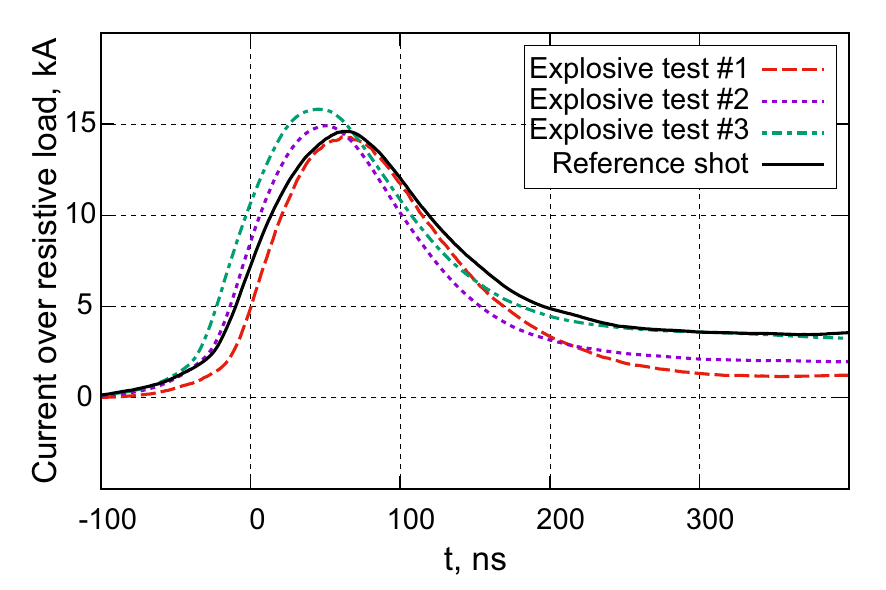}\\
	\caption{Comparison of currents over resistive load with resistance $R_d=37~\Omega$ for the reference shot and for explosive tests}
	\label{fig:Id_compare}
\end{figure}

For pre-final testing it is necessary to validate voltage and current pulses to be supplied to vacuum tube. 
Resistive load (water-filled resistor)  is used for both laboratory test (as a reference shot) and explosive test instead of vacuum tube. In reference shot the voltage and current delivered to EWA correspond to those in the secondary transformer winding in the explosive test.
Example of comparison of currents  over resistive load with resistance $R_d=37~\Omega$ for the reference shot and for explosive tests with FCG of type \#2 at crowbar closing current $I_{CB}=3.2$~kA with EWA comprising 28 straight wires of 750mm length is shown in Fig.\ref{fig:Id_compare}.

The next step is laboratory testing of the vacuum tube with all the settings used for the reference shot - this test gives understanding of system output expected in the full-scale explosive system test. The key measured parameter is the time domain waveform of the electric field strength detected a by two receiving antennas Geozondas 1-4.5\,GHz at 10~m distance on the axis of radiation pattern main lobe  and recorded by
Tektronix oscilloscopes TDS7704 and TDS7354. 
The same parameter is used to  evaluate system operation at explosive test, along with current derivative in FCG circut, FCG and diode currents.
All the measuring
channels are synchronized.

\begin{figure}[!h]
	\centering 
	\includegraphics[width=8 cm]{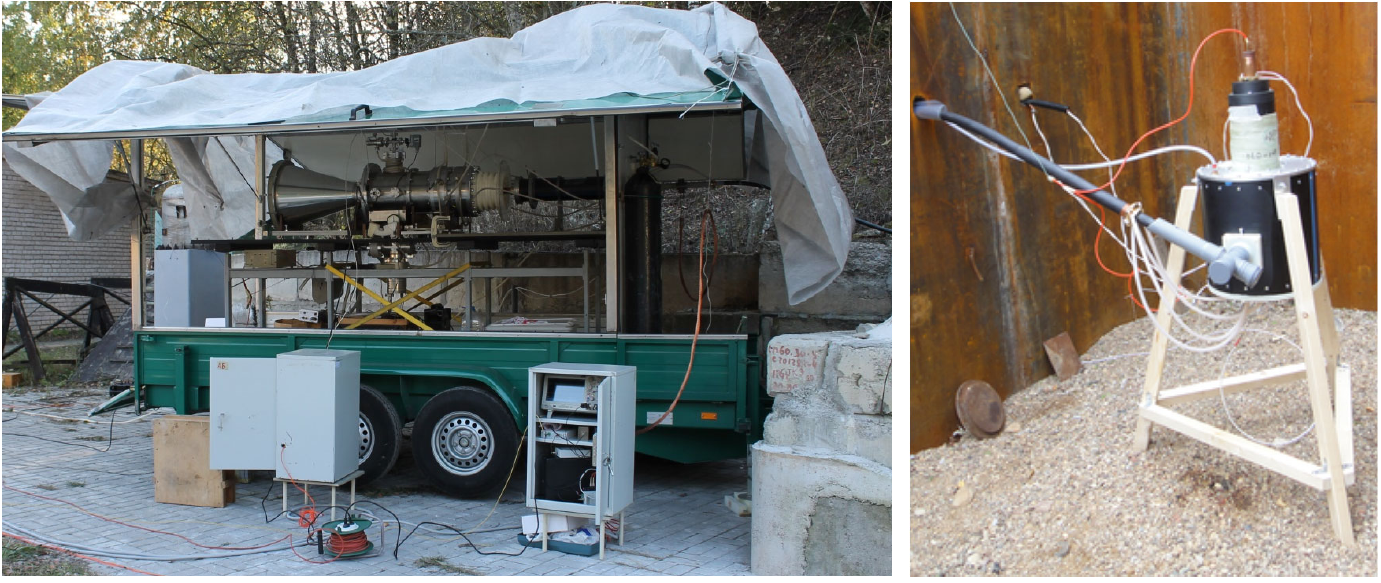}\\
	\caption{Explosively driven axial vircator: vacuum tube with EWA in a car trailer (left), FCG with transformer (right)}
	\label{fig:axial_test}
\end{figure}

\begin{figure}[!h]
	\centering 
	\includegraphics[width=8 cm]{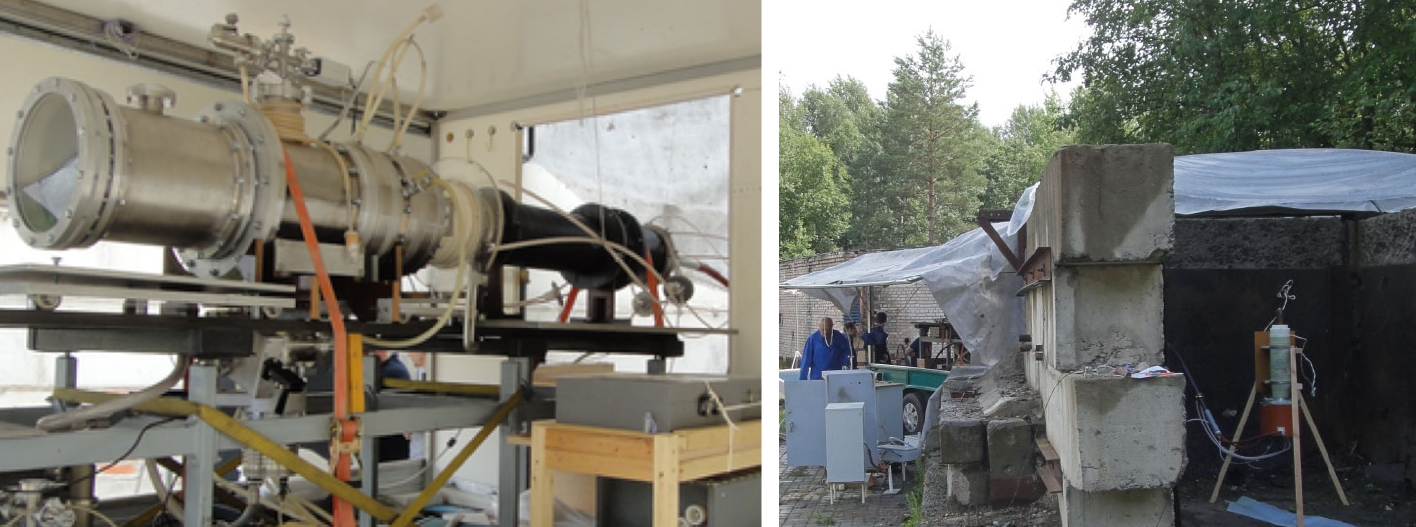}\\
	\caption{Explosively driven reflex triode: vacuum tube with EWA in a car trailer (left), FCG with transformer (right)	}
	\label{fig:reflex_test}
\end{figure}

Explosive tests were carried for both axial vircator (see Fig.\ref{fig:axial_test}) and reflex triode (see Fig.\ref{fig:reflex_test}).
Two design options of transformer were used; oil and gas insulation were tested.

\begin{figure}[!h]
	\centering 
	\includegraphics[width=8 cm]{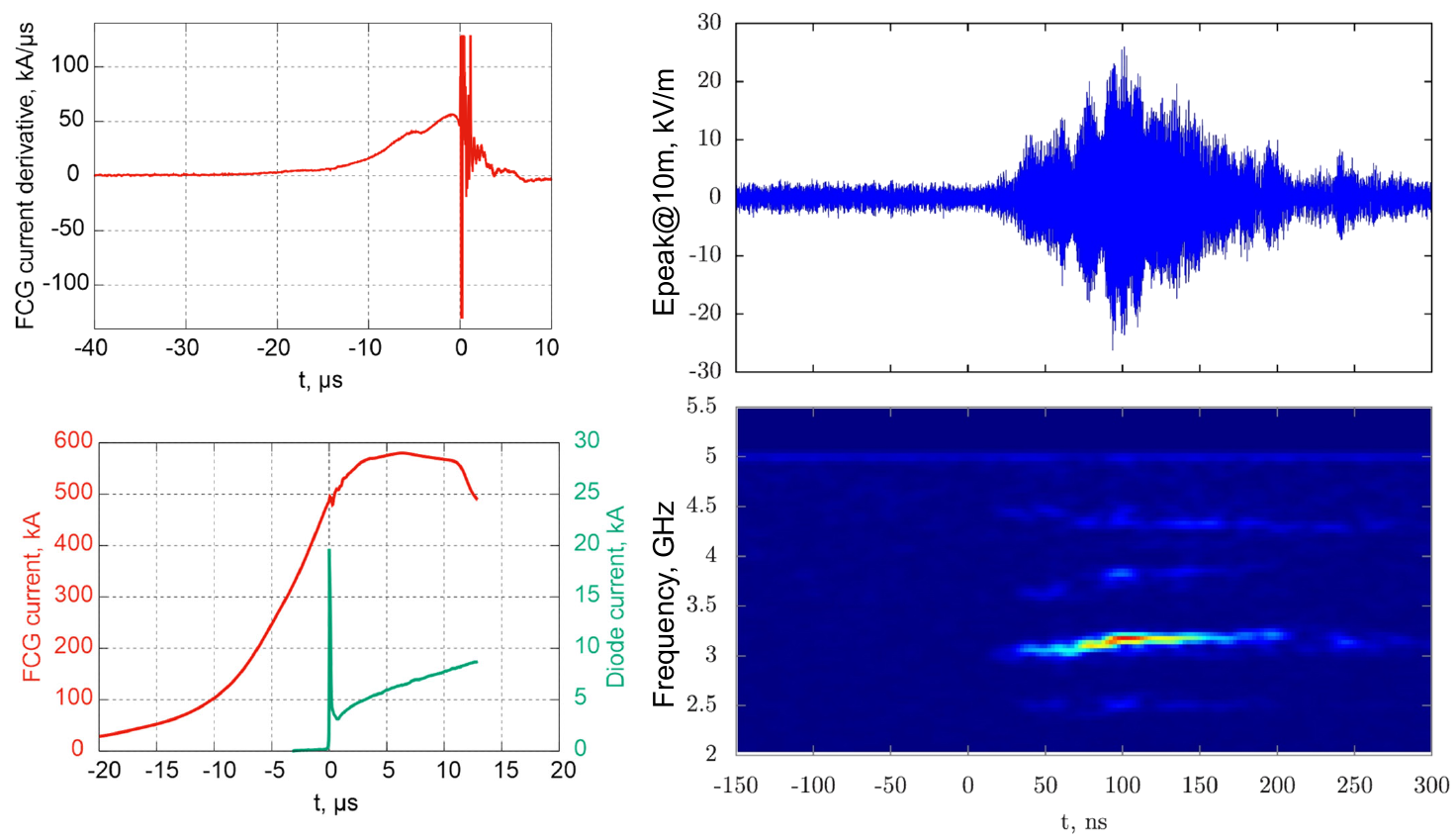}\\
	\caption{Signals recorded during test of explosively driven axial vircator}
	\label{fig:signals_axial}
\end{figure}

Axial vircator was tested with gas-insulated transformer with inductance of primary winding $L_1=290$~nH, inductance of secondary winding $L_2=7.96~\mu$H, mutual inductance $M_{12}=1.36~\mu$H, total inductance in secondary circuit 13~$\mu$H driven by FCG of type \#3 at crowbar closing current $I_{CB}=2.1$~kA, 
with gas-insulated EWA comprising 36 straight wires of 750mm length. Signals recorded during explosive test are presented in Fig.\ref{fig:signals_axial}.

\begin{figure}[!h]
	\centering 
	\includegraphics[width=8 cm]{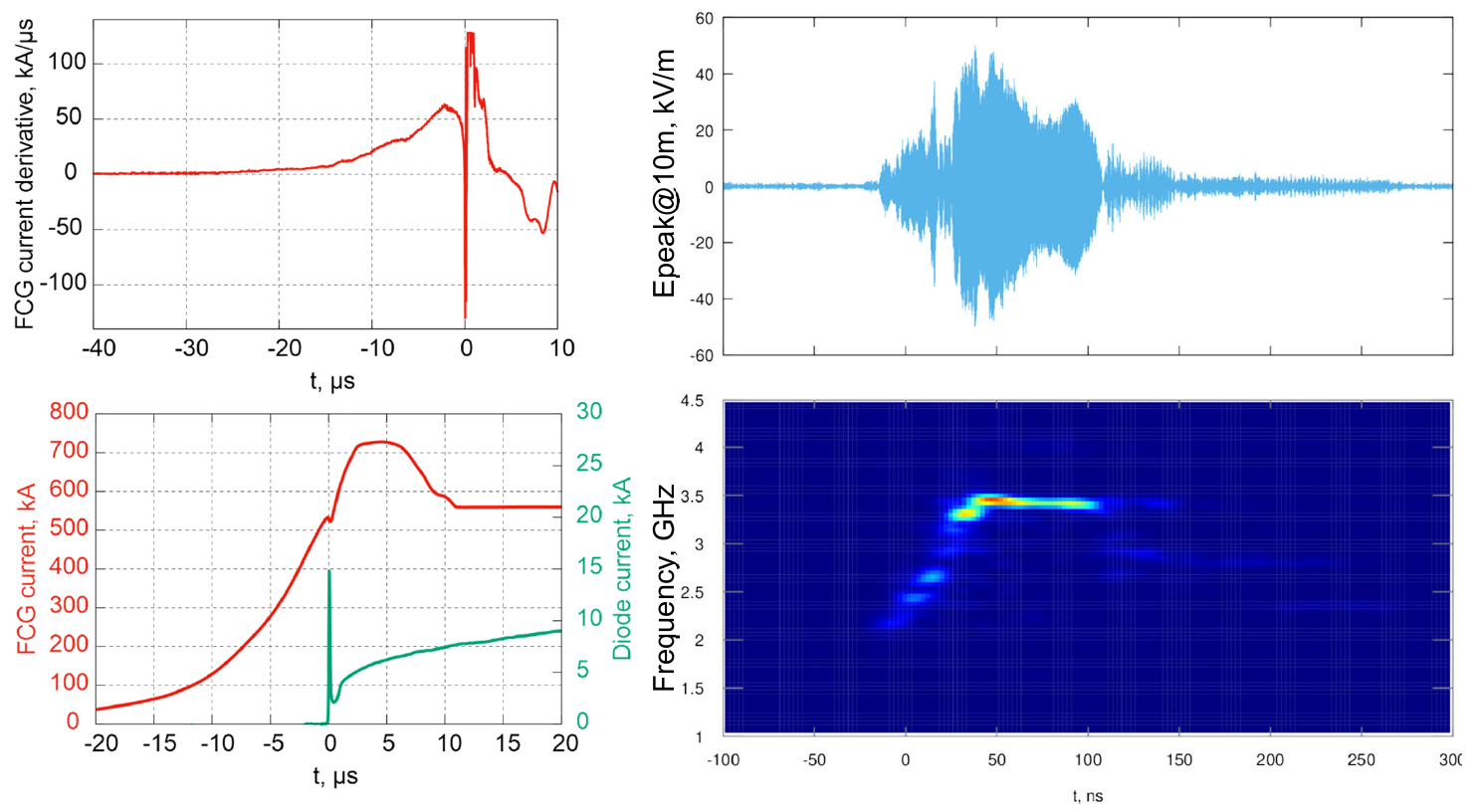}\\
	\caption{Signals recorded during test of explosively driven reflex triode}
	\label{fig:signals_reflex}
\end{figure}

Reflex triode was tested with oil-insulated transformer with inductance of primary winding $L_1=320$~nH, inductance of secondary winding $L_2=8.87~\mu$H, mutual inductance $M_{12}=1.29~\mu$H, total inductance in secondary circuit 15.6~$\mu$H driven by FCG of type \#2 at crowbar closing current $I_{CB}=3.2$~kA, 
with gas-insulated EWA comprising 28 straight wires of 750mm length. Signals recorded during explosive test are presented in Fig.\ref{fig:signals_reflex}.

\section{Conclusion}

Successive subsystem level testing enables futher perfect matching by precise defining settings and parameters for each subsystem. 
Proposed and tested matching approach ensures optimal operation of the system as a whole and enables to obtain the expected output.
All specific features of test layout (connections) should be considered at subsystem level testing: otherwise changes of layout could require subsystem re-design. 
Output parameters of an explosive subsystem could have high spread in values (10\% is a reasonable expectation),  therefore all other subsystems should have low sensitivity to such deviations.
 
System parameters are very flexible, when properly designed, matching of pulsed power source and load can be done by simple change of length and number of wires in EWA.

\end{document}